\documentclass[twocolumn]{aastex6}
\usepackage{natbib}
\usepackage{graphicx}
\usepackage{epstopdf}
\usepackage{color}
\usepackage{listings}

\newcommand{\flux}{erg~s$^{-1}$~cm$^{-2}$~}
\newcommand{\egs}{erg~s$^{-1}$~}
\newcommand{\kms}{km s$^{-1}$} 
\newcommand{\lya}{Ly$\alpha$}
\newcommand{\ha}{H$\alpha$}
\newcommand{\hb}{H$\beta$}

\newcommand{\nv}{NV$\lambda$1240}

\newcommand{\civ}{CIV$\lambda$1548,1551}
\newcommand{\heii}{HeII$\lambda$1640}
\newcommand{\uvoiii}{OIII]$\lambda$1661,1666}
\newcommand{\ciii}{CIII]$\lambda$1907,1909}

\def\arcsec{\hbox{$^{\prime\prime}$}}

\begin{document}

\title{Lyman Alpha Galaxies in the Epoch of Reionization (LAGER): Spectroscopic Confirmation of Two Redshift$\sim7.0$ Galaxies}

\author{Huan Yang\altaffilmark{1}, Leopoldo Infante\altaffilmark{1}, James E. Rhoads\altaffilmark{2}, Weida Hu\altaffilmark{3}, Zhenya Zheng\altaffilmark{4}, Sangeeta Malhotra\altaffilmark{2}, Junxian Wang\altaffilmark{3}, Felipe Barrientos\altaffilmark{5}, Wenyong Kang\altaffilmark{3}, Chunyan Jiang\altaffilmark{4}}

\altaffiltext{1}{Las Campanas Observatory, Carnegie Institution for Science; hyang@carnegiescience.edu}
\altaffiltext{2}{NASA Goddard Space Flight Center}
\altaffiltext{3}{CAS Key Laboratory for Research in Galaxies and Cosmology, Department of Astronomy, University of Science and Technology of China}
\altaffiltext{4}{CAS Key Laboratory for Research in Galaxies and Cosmology, Shanghai Astronomical Observatory, Shanghai 200030, China}
\altaffiltext{5}{Instituto de Astrofisica, Pontificia Universidad Catolica de Chile}

\begin{abstract}
We spectroscopically confirmed two narrow-band selected redshift 7.0 \lya\ galaxies and studied their restframe UV spectra. The \lya\ and other UV nebular lines are very useful to confirm the galactic redshifts and diagnose the different mechanisms driving the ionizing emission. 
We observed two narrowband-selected $z=7.0$ \lya\ candidates in the LAGER Chandra Deep Field South (CDFS) field with IMACS at Magellan telescope and confirmed they are \lya\ emitters at $z=6.924$ and $6.931$. In one galaxy, we also obtained deep NIR spectroscopy, which yields non-detections of the high-ionization UV nebular lines. We measured upper-limits of the ratios of CIV$\lambda$1548/\lya, \heii/\lya, OIII]$\lambda$1660/\lya, and CIII]$\lambda$1909/\lya\ from the NIR spectra. These upper-limits imply that the ionizing emission in this galaxy is dominated by normal star formation instead of AGN.

\end{abstract}

\section{Introduction}
Reionization of hydrogen in the intergalactic medium (IGM) was a landmark in structure formation of the early universe. Resonant scattering of \lya\ photons is sensitive to neutral hydrogen in the IGM, making Ly$\alpha$ emitters a sensitive, practical, and powerful probe of the central phase of reionization (Dijkstra et al. 2014).  
Previous studies have found thousands of \lya\ emitting galaxies (LAEs) at $z\gtrsim$ 5.7 and  6.6. The \lya\ luminosity function observed at $z\approx 6.6$ shows modest evolution from those at $z\approx 5.7$ (e.g. Malhotra \& Rhoads 2004; Hu et al. 2010; Ouchi et al. 2010; Kashikawa et al. 2011; Konno et al. 2018). The current estimate of the neutral hydrogen fraction of the IGM is $x_{HI}=0.3\pm0.2$ at $z\approx6.6$ based on \lya\ luminosity function (Konno et al. 2018). At $z \gtrsim 7.0$, \lya\ searches (Tilvi et al. 2010; Konno et al. 2014; Zheng et al. 2017; Ota et al. 2017; Itoh et al. 2018) have found a few tens LAEs and shown a significant drop of the \lya\ luminosity function relative to the one at $z\approx 6.6$. 

In the ``LAGER" project (Lyman-Alpha Galaxies in the Epoch of Reionization)
we search $z\approx 7$ LAEs by taking deep narrowband images with CTIO/DECam. In the first result, we found 23 $z\approx 7$ \lya\ emitters and showed that the $z\approx 7$ \lya\ luminosity function in the COSMOS deep field has an excess of the brightest source (Zheng et al. 2017). We have confirmed these bright LAEs using spectroscopy from Magellan/IMACS (Hu et al. 2017). 

With many $z\approx 7$ LAEs being found, we can start to understand their roles in  reionization. Do these LAEs have strong Lyman continuum emission? What mechanisms drive their galactic properties? In a few $z\sim6-9$ galaxies, observations have found high ionization UV emission lines, such as \civ, \heii, \nv, and \uvoiii\ (Stark et al. 2015a; Sobral et al. 2015; Mainali et al. 2017; Laporte et al. 2018), and suggested the presence of very hot metal poor stars, active galactic nuclei (AGN),  metal free Population III stars, or directly collapsing black hole. However, observations of a few other bright LAEs at $z\sim6-7$ have found non-detections of strong UV nebular lines with deep near-infrared (NIR) spectroscopy (Shibuya et al. 2017; Mainali et al. 2018).  

To confirm our photometric selected $z\approx 7$ LAEs and study their properties, we started a program to take their spectra with Magellan telescopes. In this article, we report the optical and NIR observations of $z\approx 7$ LAEs found in the LAGER CDFS field.

\begin{figure*}[htb]
\centering
\includegraphics[width=0.95\textwidth]{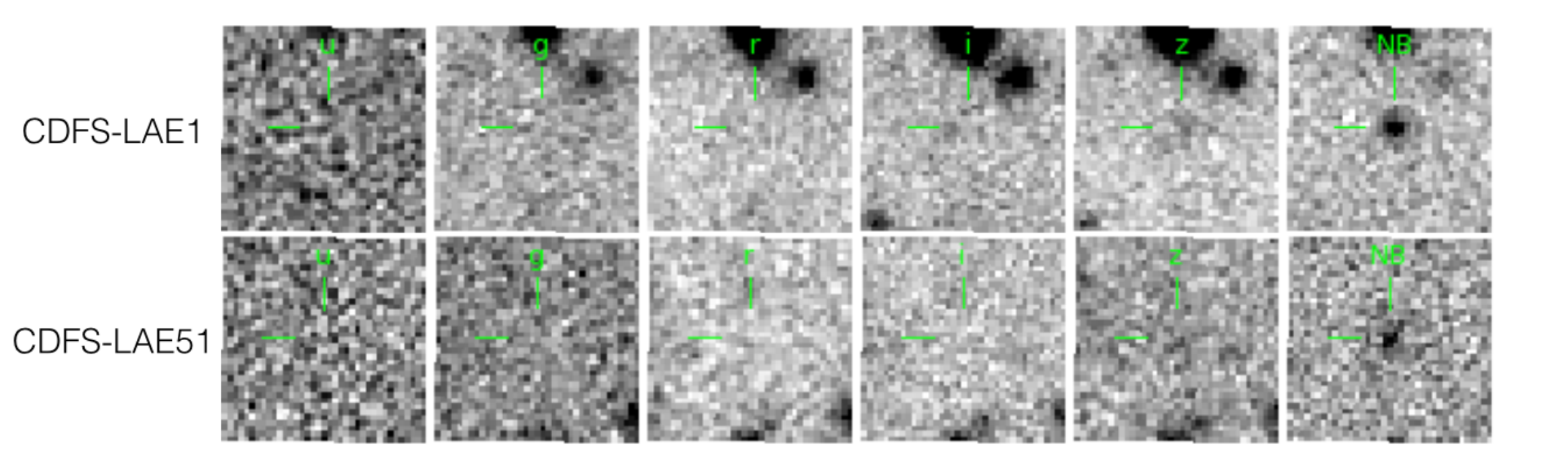}
\caption{From left to right, the DECam $u$, $g$, $r$, $i$, $z$ broadbands and narrowband NB964 images of CDFS-LAE1 and CDFS-LAE51.}
\end{figure*}

\begin{figure*}[htb]
\centering
\includegraphics[width=0.95\textwidth]{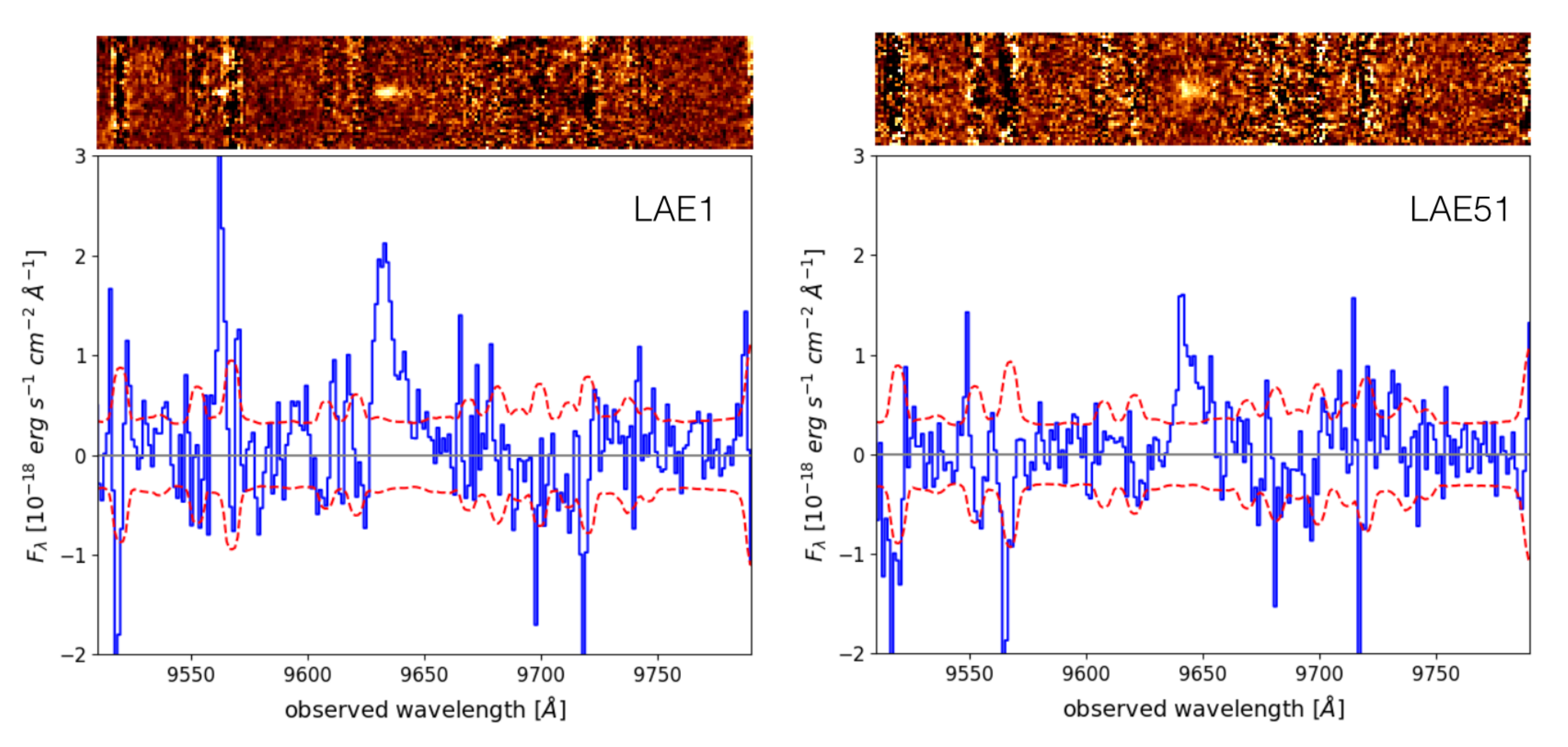}
\caption{IMACS 2D and 1D spectra of the two observed LAEs CDFS LAE1 and LAE51 around the respective line detections. The blue solid line is the spectra and the red dashed line is the 1$\sigma$ error.}
\end{figure*}

\begin{figure*}[htb]
\centering
\includegraphics[width=\textwidth]{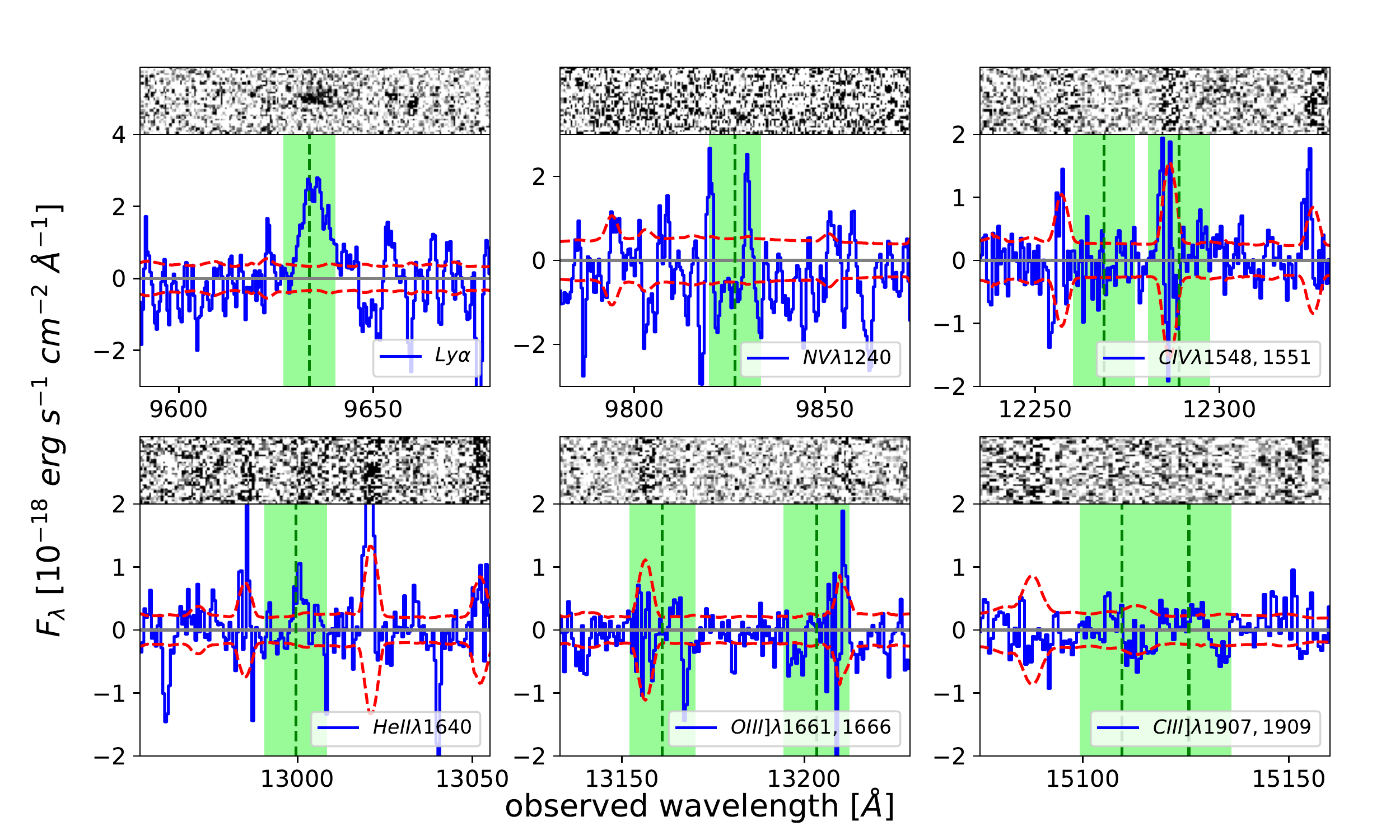}
\caption{FIRE 2D and 1D spectra of CDFS-LAE1 at the wavelength of \lya, \nv, \civ, \heii, \uvoiii, \ciii\ lines. The blue solid line is the spectra and the red dashed line is the 1$\sigma$ error. The green vertical dashed line shows the expected line positions assuming redshift = 6.9245 which is derived from the peak of observed \lya\ profile. The green shades show $\pm$200 \kms\ wavelength regions around the expected line position. No UV metal lines are detected at S/N$>5$.}
\end{figure*}

\section{Observations and Data Reduction}
In the ``LAGER" project we take deep narrowband images with CTIO/DECam (a field-of-view of 3 deg$^{2}$).  We use an optimally designed narrowband NB964 filter (FWHM $\sim$ 90\AA) to identify \lya\ lines at $z\approx 7.0$, where they fall in a dark wavelength region between strong night-sky OH lines. 
We have gathered 44 hours narrowband exposure in the COSMOS field and 34 hours narrowband exposure in the CDFS field. 
The selection of LAE candidates is shown in Hu et al. (2019). In our first two observation runs, we obtained optical and NIR spectra of two LAE candidates in the CDFS field with IMACS (Dressler et al. 2011) and FIRE (Simcoe et al. 2013) respectively at Magellan telescope. Below we describe the details of observations and data reduction. 

\subsection{Narrowband and broadband imaging}

In CDFS field, our narrowband and broadband images reach 5$\sigma$ limiting AB magnitudes of 25.03, 23.3, 27.5, 27.5, 27.3, 27.0 (2\arcsec diameter aperture) in the {\it NB964, u, g, r, i, z } bands. 
We selected LAE candidates with narrowband color excess $NB964 - z <-1.5$ and non-detections in the {\it u, g, r, i} bands (candidate selection is presented in Hu et al. 2018). 
Figure 1 shows the imaging cutouts of the two observed LAEs -- CDFS-LAE1 (RA=03:33:05.95 DEC=-28:24:59.10 J2000) and CDFS-LAE51 (RA=03:32:51.68 DEC=-28:32:42.51 J2000). 

\subsection{Magellan/IMACS spectroscopy}
We observed 2 LAEs using IMACS on the 6.5m Magellan I Baade Telescope on 2017 October 20$-$21. We used the IMACS f/2 camera with 300-line red-blazed grism and multi-slits mask. The mask has 1\arcsec\ slit width and covers the two bright z$\approx$7.0 LAEs, dropout selected galaxies, other emission line galaxies at intermediate redshifts, flux calibration stars, and alignment stars. On Oct 20, the sky was clear and seeing $\sim0.6-0.8\arcsec$\ and we observed the mask in the second-half night for 220 minutes (12 $\times$ 20 minutes per exposure). On Oct 21, the weather was partly cloudy with passing clouds and we got 60 minutes observations under clear sky condition.

The spectra coverage is 6000$-$11000 \AA. The spectra resolution is about 194 \kms\ at 9640\AA.

We reduced the IMACS data with the pipeline COSMOS2 (Oemler et al. 2017). 
Following the standard pipeline procedures, we get the stacked wavelength-calibrated 2D spectra. 
We stacked 280 minutes exposure into one final 2D spectra. 
The 1$\sigma$ flux limit at 9640\AA\ is 1.8$\times10^{-18}$ \flux. 
Then we extract the spectra within an aperture of 1.2\arcsec\ (box extraction).   
We fit a blackbody model to the spectra of one flux-calibration star on the mask to get the sensitivity function and applied it to the 1D spectra of the LAEs.

\subsection{Magellan/FIRE spectroscopy}
We observed these two LAEs using Magellan/FIRE in Echelle mode on 2017 December 26$-$27. 
On Dec 26, the seeing was 0.6$-$0.7 \arcsec\ and we observed CDFS-LAE-1 for 200 minutes with the 0.75\arcsec\ slit. One Dec 27, the seeing was 1.0$-$1.2\arcsec\ and we observed CDFS-LAE-51 for 220 minutes with 0.75\arcsec\ slit for the first one hour and 1\arcsec\ slit for the remaining time. A telluric star was observed every 1$-$2 hours throughout the night. 

We reduced the FIRE data with the FIREHOSE pipeline. After the 2D sky subtraction, the 1D LAE spectra was extracted in each frame using the trace and spatial profile of the telluric star as reference. Then the 1D spectra were stacked together. The 1D spectra covers wavelength range 8500$-$21000 \AA. The 1D spectra has 1$\sigma$ flux limit of 1.4$\times10^{-18}$ \flux\ at 9640\AA.

\section{Results}

\subsection{LAE confirmations}
In the optical and NIR spectra, one line at wavelength $\sim$ 9640\AA\ is clearly detected for both LAEs. Figure 2 shows the IMACS 2D and 1D spectra of these two emission lines. These two lines are detected at S/N=12 and 9 respectively.
These two lines could be \lya\ lines at $z\sim7.0$ or lower redshift emission lines. 
We ruled out the possibility of low redshift lines because 1) they are not detected in the deep $g-r-i$ bands images reaching 5$\sigma$ limiting magnitudes of 27.5 mag; 2) strong optical lines, such as [OIII]$\lambda$4959,5007, [OII]$\lambda$3727, \ha, and \hb\ have companion lines that were covered in the observed spectra but not detected. For example, if the line is [OIII]$\lambda$5007 (or [OIII]$\lambda$4959), we expect to see the [OIII]$\lambda$4959 (or [OIII]$\lambda$5007) in the IMACS spectra and the \ha\ line in the near-infrared spectra; if the line is [OII]$\lambda$3727, we expect to see the [OIII]$\lambda$4959,5007, \ha\ and \hb\ lines in the near-infrared spectra. Yet those companion lines are not detected. 3) the line profiles are asymmetric with red wings, which are typical of high-redshift \lya\ emission lines.
Therefore we conclude these two lines are \lya\ lines. The \lya\ redshifts for LAE1 and LAE51 are 6.9245 and 6.931.

We measure the \lya\ line fluxes by integrating the spectra range of $-200 \sim 450$ \kms\ around the \lya\ line in the IMACS spectra. The \lya\ line fluxes are 21.8 $\pm$ 1.8 $\times10^{-18}$ \flux\ for LAE1 and 15.6 $\pm$ 1.8 $\times10^{-18}$ \flux\ for LAE51 (table 1). The full width at half maximum (FWHM, corrected for instrumental resolution) of \lya\ line for LAE1 and LAE51 are 229 and 318 \kms. The \lya\ line luminosities are $1.21\times10^{43}$ \egs\ for LAE1 and $0.87\times10^{43}$ \egs\ for LAE51. Assuming 10\% \lya\ escape fraction and \lya/\ha=8.7 in Case-B recombination and the prescriptions for calculating star-formation rates in Kennicutt \& Evans (2012), the corresponding star-formation rates are 75 $M_{\odot}~yr^{-1}$ for LAE1 and 54 $M_{\odot}~yr^{-1}$ for LAE51. 

Notice that in the spectra of LAE1 there is another very narrow spike at 9562\AA\ which overlaps with sky line. Its wavelength is covered in the FIRE spectra but no signal is detected at 9562\AA. Therefore we discard it as an artifact from imperfect sky subtraction.

\begin{deluxetable}{ lllr }
\tablecaption{Emission lines measurements}
\centering
\tabletypesize{\scriptsize}
\tablehead{\colhead{Object} & \colhead{Redshift} & \colhead{Line} & \colhead{Flux ($10^{-18}$} \\
\colhead{} & \colhead{} & \colhead{} & \colhead{\flux)}  }
\startdata
LAE1 & 6.9245 & \lya\ (IMACS)  &  21.8 $\pm$ 1.8\\
 & & \lya\ (FIRE)  &  20.6 $\pm$ 1.4\\
 & & NV1240       &   $<$6.4 \\
 & & CIV1548.2   &  $<$1.4 \\
 & & CIV1550.8   & $<$5.0 \\
 & & HeII1640.4  &  $<$1.7 \\
 & & OIII]1660.8  &  $<$2.8 \\
 & & OIII]1666.2  &  $<$1.9 \\
 & & CIII]1907.0  &  $<$1.7 \\
 & & CIII]1909.0  &  $<$1.2 \\
\hline
LAE51 & 6.931 & \lya\ (IMACS) & 15.6 $\pm$ 1.8 \\
\enddata
\tablecomments{The non-detections of UV nebular Lines are 2$\sigma$ upperlimits. The errors of \lya\ line fluxes are 1$\sigma$ errors.}
\end{deluxetable}

\begin{figure*}[htb]
\centering
\includegraphics[width=0.85\textwidth]{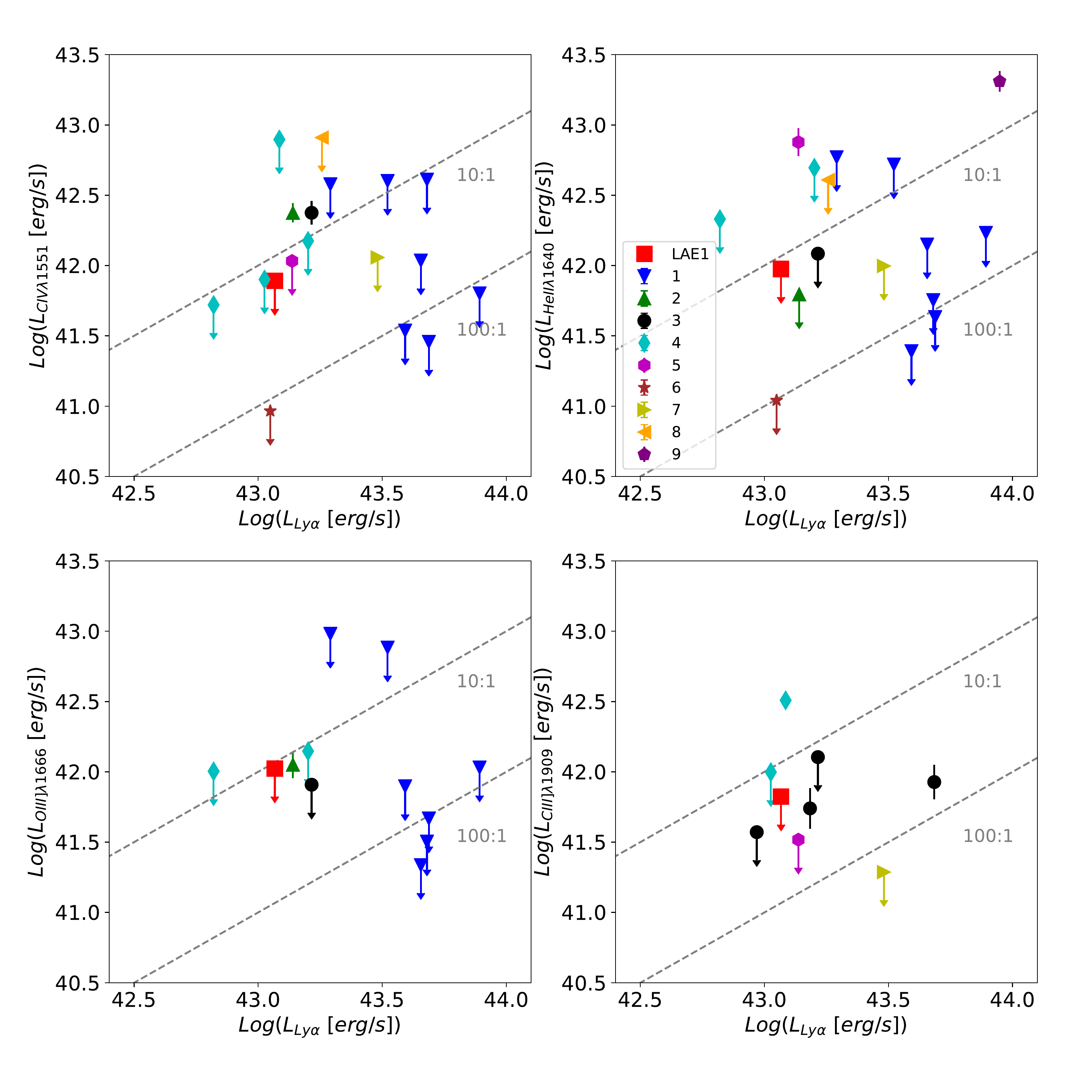}
\caption{Comparisons of \lya\ and CIV$\lambda$1551, \heii, OIII]$\lambda$1666, and CIII]$\lambda$1909 for LAE1 and other $z=5.6\sim8.6$ galaxies in the literature. Points without downward arrows are detections (at $\sim3-5 \sigma$ level). Points with downward arrows are $2\sigma$ upper limits. We assume the two single lines have equal fluxes for the doublets of CIV$\lambda$1548,1551, OIII]$\lambda$1661,1666, and CIII]$\lambda$1907,1909. When upper-limits of the two single lines of doublets were reported in the literature, we used the smaller upper-limits. The dashed lines show \lya\ to UV nebular line ratios of 10:1 and 100:1. The numbers in the legend represent the following references: (1) Shibuya et al. 2017; (2) Mainali et al. 2017; (3) Stark et al. 2015a,b; (4) Mainali et al. 2018; Stark et al. 2017 (5) Laporte et al. 2017; (6) Kashikawa et al. 2012; (7) Zabl et al. 2015; (8) Nagao et al. 2005; (9) Sobral et al. 2015. Note that a few galaxies in references (2), (3), (4) are gravitationally-lensed, and the plots show their uncorrected observed lines luminosities.}
\end{figure*}

\subsection{UV nebular lines}
Although \lya\ is the only detected line in the spectra, we can put limits on the other UV nebular lines from the FIRE spectra. 
In the FIRE spectra of CDFS-LAE1, the \lya\ line is clearly detected with line flux of 20.6 $\pm$ 1.4 $\times10^{-18}$ \flux\ and intrinsic FWHM of 218 \kms. We focus on this object in the following analysis. In the FIRE spectra of CDFS-LAE51, the \lya\ line is not detected with a 3$\sigma$ flux upper-limit of 9.0 $\times10^{-18}$ \flux, which is slightly lower than the line flux in IMACS spectra. This could be caused by larger slit loss at poor seeing ($\sim$ 1.0$-$1.2\arcsec) and possibly larger blind-offset errors during the FIRE observation.  

In figure 3, we show the spectra of CDFS-LAE1 at the wavelength of \lya, \nv, \civ, \heii, \uvoiii, and \ciii\ lines. The expected line positions are calculated assuming a redshift of 6.9245. For the non-detected UV nebular lines, we assume a line width of 200 \kms\ and calculate the 2$\sigma$ upper-limits (table 1).  

In CDFS-LAE1, the 2$\sigma$ upper-limits of the flux ratios of CIV$\lambda$1548/\lya, \heii/\lya, OIII]$\lambda$1660/\lya, and CIII]$\lambda$1909/\lya\ are about 0.06$-$0.1. 
There are two spikes near the position of \nv\ line. Both spikes are very narrow with FWHM $\sim$ 20 \kms, and they are not shown in the rectified 2D spectra. Therefore they were discarded as noise spikes. There are also marginal signal in the position of \heii\ line, but the current data is inconclusive.

We considered the contributions of active galactic nucleus in these galaxies. Because the \lya\ line is very narrow with FWHM $\sim$ 220 \kms, we compared it to the narrow lines of AGNs. The narrow CIV$\lambda$1548 line to \lya\ ratios of narrow-line AGNs at intermediate redshifts are about 0.1$-$0.2 (Alexandroff et al. 2013). The CIV$\lambda$1548 to \lya\ ratio in CDFS-LAE1 is smaller than 0.07. Therefore the contribution from an AGN is small and the H ionizing emission is dominated by normal star formation instead of AGN.

We also compared the ratios in LAE1 with similar measurements for \lya\ galaxies at $z=5.6-8.6$ in the literature in figure 4. LAE1 has a \lya\ luminosity of $1.21\times10^{43}$ \egs, which is relatively lower than the other galaxies. The line ratios of LAE1 are consistent with the other measurements as shown in figure 4. Among these high redshift galaxies, except for a few detections at $3-5\sigma$ levels, most flux measurements of UV nebular lines are upper-limits at $\sim 1-10\%$ of \lya\ line flux. To diagnose the ionizing properties of LAEs, the UV spectra should ideally reach a depth of $1-5\%$ \lya\ line flux. 
Because the CIV, CIII] or HeII lines are very weak relative to \lya\ lines in these LAEs, unless \lya\ overlaps with strong sky lines, it is easier to use \lya\ emission line to measure the galactic redshift.

\section{Conclusions}

We spectroscopically confirmed two redshift $\sim$ 7 \lya\ galaxies (LAE1 and LAE51) found in the LAGER survey. In LAE1, our deep NIR spectroscopy yields non-detections of the high-ionization UV nebular lines. We derived strong upper-limits of the ratios of CIV$\lambda$1548/\lya, \heii/\lya, OIII]$\lambda$1660/\lya, and CIII]$\lambda$1909/\lya. Because its CIV$\lambda$1548 line to \lya\ ratio is smaller than the typical ratio in AGNs,  the ionizing emission in LAE1 is dominated by normal star formation instead of AGN.

\acknowledgements
The observations discussed here were taken with the IMACS and FIRE instruments on the Magellan/Baade 6.5m telescope at Las Campanas Observatory, Chile.  
W.D.H, J.X.W. and W.Y.K. thanks support from National Science Foundation of China (grants No. 11421303 \& 11890693) and CAS Frontier Science Key Research Program (QYZDJ- SSW-SLH006), and National Basic Research Program of China (973 program, grant No. 2015CB857005). Z.Y.Z. is sponsored by Shanghai Pujiang Program, the National Science Foundation of China (11773051), the China-Chile Joint Research Fund (CCJRF No. 1503) and the CAS Pioneer Hundred Talents Program. L.F.B was supported by Anillo ACT-1417.

\end{document}